\documentclass[pre,aps,twocolumn,showpacs,floatfix,10pt,superscriptaddress]{revtex4}
\usepackage{graphicx}
\usepackage{color}
\usepackage{amsmath, amsthm, amssymb}
\newcommand{\be}{\begin{equation}}
\newcommand{\ee}{\end{equation}}
\newcommand{\bea}{\begin{eqnarray}}
\newcommand{\eea}{\end{eqnarray}}

\begin{document}

\title{The nematic-disordered phase transition in systems of long rigid rods 
on two dimensional lattices}
\author{Joyjit Kundu}
\email{joyjit@imsc.res.in}
\affiliation{The Institute of Mathematical Sciences, C.I.T. Campus,
Taramani, Chennai 600113, India}
\author{R. Rajesh}
\email{rrajesh@imsc.res.in}
\affiliation{The Institute of Mathematical Sciences, C.I.T. Campus,
Taramani, Chennai 600113, India}
\author{Deepak Dhar}
\email{ddhar@theory.tifr.res.in}
\affiliation{Department of Theoretical Physics, Tata Institute of
Fundamental
Research, Homi Bhabha Road, Mumbai 400005, India}
\author{J\"{u}rgen F. Stilck}
\email{jstilck@if.uff.br}
\affiliation{Instituto de F\'{i}sica and National Institute of Science
and
Technology for Complex Systems, Universidade Federal Fluminense, Av.
Litor\^anea s/n, 24210-346 Niter\'oi, RJ, Brazil}

\date{\today}

\begin{abstract}
We study the phase transition from a nematic phase to a high-density 
disordered phase in systems of long rigid rods of length $k$ on the 
square and triangular lattices. We use an efficient Monte Carlo scheme 
that partly overcomes the problem of very large relaxation times of 
nearly jammed configurations.  The existence of a continuous transition 
is observed on both lattices for $k=7$.  We study 
correlations in the high-density disordered phase, and we find evidence 
of a crossover length scale $\xi^* \gtrsim 1400$, on the square lattice.  
For distances smaller than $\xi^*$, correlations appear to decay 
algebraically.  Our best estimates of the critical exponents differ from 
those of the Ising model, but we cannot rule out a crossover to Ising 
universality class at length scales $\gg \xi^*$. On the triangular 
lattice, the critical exponents are consistent with those of the two 
dimensional three-state Potts universality class.
\end{abstract}

\pacs{64.60.Cn,64.70.mf,05.50.+q}

\maketitle

\section{\label{sec:intro}Introduction}

The study of the ordering transition in a system of long rigid rods in 
solution with only excluded volume interaction has a long history, 
starting with Onsager's proof that beyond a critical density, a solution 
of thin cylindrical rods would undergo a transition to an 
orientationally ordered 
state~\cite{onsager1949,flory1956a,zwanzig1963,vroege1992}. In two 
dimensional continuum space, when the rods may orient in any direction, 
the continuous rotational symmetry remains unbroken at any density. 
However, the system undergoes a Kosterlitz-Thouless type transition from 
a low-density phase with exponential decay of orientational correlations 
to a high-density phase with a power law decay 
\cite{straley1971,frenkel1985,khandkar2005,vink2009}.

In this paper, we study the problem when the underlying space is 
discrete: the square or the triangular lattice. Straight rods occupying 
$k$ consecutive sites along any one lattice direction will be called 
$k$-mers. For dimers ($k=2$), it has been shown rigorously that the 
system remains in the isotropic phase at all packing densities 
\cite{lieb1972}. A system of dimers with additional interactions, 
either attractive, favoring parallel alignment, or 
repulsive, disallowing nearest neighbor occupation, can have ordered 
phases \cite{dickman2012}. For $k>2$, the existence of a phase 
transition, with only hard-core interactions, remained unsettled for a 
long time \cite{degennesBook}. Ghosh and Dhar recently argued that 
$k$-mers on a square lattice, for $k \geq k_{min}$, would undergo two 
phase transitions, and the nematic phase would exist for only an 
intermediate range of densities $\rho_1^* < \rho < \rho_2^*$ 
\cite{ghosh2007}. Similar behavior is expected in higher dimensions.  In 
two dimensions, numerical studies have shown that $k_{min}=7$ 
\cite{ghosh2007}. The existence of the nematic phase, and hence the 
first transition from the isotropic to nematic phase, has been proved 
rigorously \cite{giuliani2012}. This transition has also been studied in 
detail through Monte Carlo simulations \cite{fernandez2008a, 
fernandez2008b, fernandez2008c,linares2008}. On the square lattice, the 
transition is numerically found to be in the Ising (equivalently the 
liquid-gas) universality class \cite{fernandez2008a, fischer2009}, and 
on the triangular lattices, it is in the $q=3$ Potts model universality 
class \cite{fernandez2008a,fernandez2008b}.

In this paper, we investigate whether the high-density disordered phase 
is a reentrant low-density disordered phase, or a new qualitatively 
distinct phase.  To distinguish between these two phases without nematic 
order, we will refer to the first as low-density disordered (LDD) phase 
and the second as high-density disordered (HDD) phase in the remainder 
of the paper.

The second transition at $\rho_2^*$ from the nematic to the HDD phase 
has not been studied much so far. Numerical studies are difficult 
because of the large relaxation times of the nearly jammed 
configurations at high densities.  Conventional Monte Carlo algorithms 
using deposition-evaporation moves involving only addition or removal of 
single rods at a time are quite inefficient at large densities. With 
additional diffusion and rotation moves, it is possible to equilibrate 
the system \cite{linares2008,barnes2009}, but the algorithm is still not 
efficient enough to make quantitative studies of the transition, or the 
nature of the HDD phase. In Ref.~\cite{ghosh2007}, a variational 
estimate of the entropy of the nematic and the HDD phases suggests that 
$1 -\rho_2^*$ should vary as $1/k^2$ for large $k$. Linares et. al. 
estimated that $ 0.87 \leq \rho_2^* \leq 0.93$ for $k = 7$, and proposed 
an approximate functional form for the entropy as a function of the 
density \cite{linares2008}.  However, not much is known about the nature 
of transition. Recently, we showed that a Bethe-like approximation 
becomes exact on a random locally tree like layered lattice, and for the 
$4$-coordinated lattice, $k_{min}=4$. But on this lattice, the second 
transition is absent \cite{dhar2011}.

By implementing an efficient Monte Carlo algorithm, we show that, for 
$k=7$, at high densities the orientational order is absent on both 
square and triangular lattices. We investigate the nature of this HDD 
phase. Using lattices of size up to $L = 2576$, we find 
evidence of a power-law decay of orientational correlations between rods 
for distances $r \leq \xi^* \approx 1400$, where $\xi^*$ is a 
characteristic length scale of the system. Correlations appear to decay 
faster for distances $r \gtrsim \xi^*$, but we have limited data in this 
regime, and cannot rule out a power-law decay, even for $r 
\gg \xi^*$.

Regarding the critical behavior near the phase transition on the square 
and triangular lattices, for $k=7$, our results show that the transition 
is continuous and occurs for $\rho_2^*= 0.917 \pm .015$
($\mu_c=5.57\pm .02$) on the square 
lattice, and for $\rho_2^*=0.905 \pm .010$  ($\mu_c=5.14
\pm 0.05$)
on the triangular lattice, where $\mu_c$ is the critical
chemical potential.  
For comparison, $\rho_1^*\approx 0.745$ on the square lattice. On the 
square lattice, our best estimates of the effective critical exponents 
differ from the Ising universality class, with exponents $\nu = 0.90 \pm 
0.05$, $\beta/\nu = 0.22 \pm 0.07$, $\gamma/\nu = 1.56\pm 0.07$ and 
$\alpha/\nu = 0.22 \pm 0.07$. However, it appears that these are only 
effective exponents, and may be expected to crossover to 
the Ising universality class at larger length scales. On the triangular 
lattice, our estimates of critical exponents for the second transition 
are consistent with those of the three-state Potts model universality 
class ($\nu=5/6$, $\beta=1/9$). 

The plan of the paper is as follows. In Sec.~\ref{sec:model}, we define 
the model precisely, and describe the Monte Carlo algorithm used.  In 
Sec.~\ref{sec:nucleation}, we use this algorithm to show that at high 
activities, the nematic phase is unstable to creation of bubbles of HDD 
phase, and that the decay of the nematic order parameter to zero is 
well-described quantitatively by the classical nucleation theory of 
Kolmogorov-Johnson-Mehl-Avrami. In Sec.~\ref{sec:highdensity}, we study 
different properties of the HDD phase: the two point correlations, 
cluster size distributions, susceptibility, size distribution of 
structures that we call `stacks', and the formation of bound states of 
vacancies. The critical behavior near the second transition from the 
nematic phase to the HDD phase is studied in Sec.~\ref{sec:results} for 
both the square and triangular lattices, by determining the numerical 
values of the critical exponents. Section~\ref{sec:summary} summarizes 
the main results of the paper, and discuss some possible extensions.

\section{\label{sec:model}Model and the Monte Carlo algorithm}

For simplicity, we first define the model on the square lattice. 
Generalization to the triangular lattice is straightforward. Consider a 
square lattice of size $L \times L$ with periodic boundary conditions. A 
$k$-mer, can 
be either horizontal (x-mer) or vertical (y-mer). A lattice site can 
have at most one $k$-mer passing through it.  An activity $z=e^\mu$ is 
associated with each $k$-mer, where $\mu$ is the chemical potential.

The Monte-Carlo algorithm we use is defined as follows (this was 
reported earlier in a conference \cite{joyjit_dae}): given a valid 
configuration, first, all x-mers are removed without moving any of the 
y-mers. Each row now consists of sets of contiguous empty sites, 
separated from each other by sites occupied by y-mers.  The lattice is 
now reoccupied with x-mers. In the grand canonical ensemble, this can be 
done independently in each row, and the problem reduces to that of 
occupying an interval of some given length $\ell$ of a one dimensional 
lattice with k-mers with correct probabilities.

Let the grand canonical partition function of a system of hard rods on a 
one dimensional lattice of $\ell$ sites with open boundary conditions be 
denoted by $\Omega_o(z;\ell)$. The probability that the left most site 
is occupied by the left most site of a x-mer is $p_{\ell} = z 
\Omega_o(z;\ell-k)/\Omega_o(z;\ell)$. If not occupied, we consider the 
neighbor to the right and reduce the number of lattice sites by one. If 
occupied, we move to the $(k+1)^{th}$ neighbor and reduce the length of 
the interval by $k$.

The partition functions $\Omega_o(z;\ell)$ obeys the simple recursion 
relation $\Omega_o(z;\ell) = z \Omega_o(z;\ell-k) + \Omega_o(z;\ell-1)$, 
for $\ell\geq k$, and $\Omega_o(z;\ell)=1$ for $\ell=0,1,\ldots,k-1$. 
The solution of this recursion relation is 
$\Omega_o(z;\ell)=\sum_{i=1}^{k} a_i \lambda_i^\ell$, where 
$\lambda_i$'s are independent of $\ell$. The $a_i$'s are determined by 
the boundary conditions $\Omega_o(z;\ell)=1$ for $\ell=0,1,\ldots,k-1$.

With periodic boundary conditions, the recursion relations have to be 
modified. Let $\Omega_p(z;\ell)$ be the partition function of a one 
dimensional lattice of length $\ell$ with periodic boundary conditions. 
It is easy to see that $\Omega_p(z;\ell) = k z \Omega_o(z;\ell-k) + 
\Omega_o(z;\ell-1)$.  We use a list of stored values of the relevant 
probabilities $\{p_{\ell}\}$ for all $\ell=1,\ldots,L$, to reduce the 
computation time.

Following the evaporation of and re-occupation by x-mers, we repeat the 
procedure with y-mers. Keeping x-mers unmoved, all y-mers are evaporated 
and the columns are then reoccupied with y-mers. A Monte Carlo move 
corresponds to one set of evaporation and re-occupation of both x-mers 
and y-mers. It is straightforward to see that the algorithm is ergodic, 
and satisfies the detailed balance condition.

The algorithm is easily parallelizable since the evaporation and 
reoccupation of x-mers in any row (column) is independent of the other rows (columns). For 
the larger system sizes ($L > 400$), we used a parallelized version of 
the computer program. This enables us to study the critical behavior at 
the second transition for system sizes up to $L=952$, properties of the 
HDD phase away from the transition point for system sizes up to 
$L=2576$, and probe densities up to $\rho=0.995$. At these high 
densities, we ensure equilibration by checking that the long time 
behavior of the system is independent of the initial  
preparation. For this, we used two different initial conditions, one in 
which all sites are occupied by x-mers and the other in which one half 
of the lattice contains x-mers and the other half only y-mers.

\section{\label{sec:nucleation}Metastability of the nematic phase for 
large activities}

\begin{figure*} 
\begin{center} 
\includegraphics[width=2.0\columnwidth]{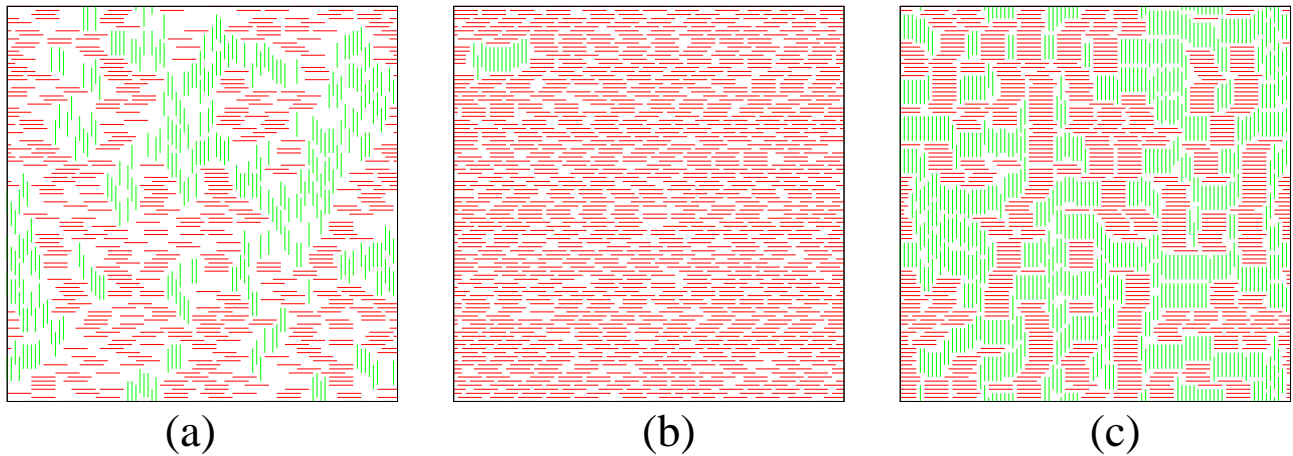} 
\caption{(Color online)
Typical configurations of the system in equilibrium
at densities (a) $\rho \approx 0.66$
($\mu=0.41$) (b) $\rho \approx 0.89$ ($\mu=4.82$), and  (c) $\rho
\approx 0.96$ ($\mu=7.60$) on a square
lattice.   Here, $k=7$ and $L=98$. 
} 
\label{fig:fig_snapshot}
\end{center}
\end{figure*}

\begin{figure} 
\includegraphics[width=6.0cm]{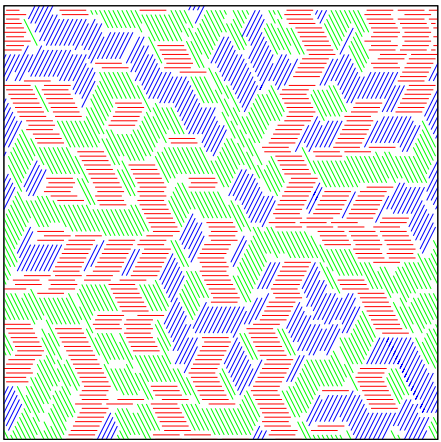} 
\caption{(Color online)
A typical configuration of the system in equilibrium 
at density $\rho \approx 0.96$ ($\mu=7.60$) on
a triangular lattice.    Here,  $k=7$ and $L=98$.
}
\label{fig:fig_snapshot_tri}
\end{figure}

We first verify that, for large activities, the nematic phase is 
unstable to the growth of the HDD phase. In 
Fig.~\ref{fig:fig_snapshot}(a)--(c), we show snapshots of the system of 
rods of length $k=7$ in equilibrium, on a square lattice at low, 
intermediate and high densities. For the high-density snapshot, the 
initial configuration had full nematic order, but the system relaxed to 
a disordered phase. A similar disordered phase is also seen for the 
triangular lattice at high densities (see 
Fig.~\ref{fig:fig_snapshot_tri}).

\begin{figure}
\includegraphics[width=\columnwidth]{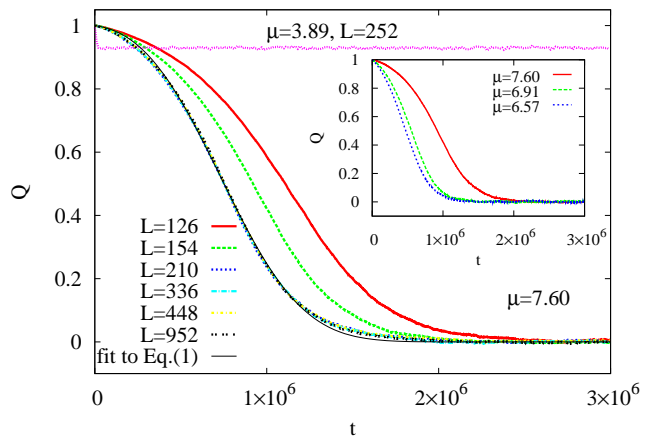}
\caption{(Color online) Decay of the order parameter $Q$ for the square lattice
as a function of time (Monte 
Carlo steps), starting from a fully ordered state for two different 
values of $\mu$: $\mu=3.89$ ($\rho \approx 0.867$), and $\mu=7.60$ 
($\rho \approx 0.957$). The best fit of the data to 
Eq.~(\ref{eq:nucleation}) with additional subleading terms is also 
shown. Inset: Data for different chemical potentials, all corresponding 
to HDD phase for $L=154$ and $k=7$. The densities corresponding to these 
values of $\mu$ are approximately $0.957$, $0.948$, $0.941$.}
\label{fig:Q_tK7L154}
\end{figure}

In Fig.~\ref{fig:Q_tK7L154}, we show the temporal evolution of the order 
parameter $Q$, defined by $Q=\langle n_h-n_v \rangle/ \langle n_h+n_v 
\rangle$, where $n_h$ and $n_v$ are the number of x-mers and y-mers 
respectively.  For all values of $\mu$, the initial configuration had 
full nematic order. For $\mu=3.89$, at large times, the system relaxes 
to an equilibrium state with a finite nematic order. However, for larger 
$\mu = 7.60$, the nematic order decreases with time to zero. 
Interestingly, we find that the average lifetime of the metastable state 
{\it decreases} with increasing system size, and saturates to a $L$ 
independent value for $L \gtrsim 200$ (see Fig.~\ref{fig:Q_tK7L154}).

Naively, faster relaxation for larger systems may appear  unexpected, but 
is easily explained  using the well-known nucleation theory of 
Kolmogorov-Johnson-Mehl-Avrami \cite{Ramos1999, rikvold1994}.  
We assume that critical droplets of the 
stable phase are created with a small uniform rate $\epsilon$ per unit 
time per unit area, and once formed, the droplet radius grows at a 
constant rate $v$. Then, the probability that any randomly chosen site 
is still not invaded by the stable phase is given by $\exp [-\epsilon 
\int_{0}^t dt' V(t')]$, where $V(t')$ is the area of the region such 
that a nucleation event within this area will reach the origin before 
time $t'$. The area $V(t')$ is given by $V(t') = \pi v^2 t'^2$ when the 
droplet is smaller than the size of the lattice. For time $t' $ greater 
than this characteristic time $t^*$, we have $V(t') =L^2$. If the 
droplet does not grow equally fast in all directions, we take suitably 
defined average over directions to define $v^2$.  Thus, we obtain
\begin{eqnarray}
Q(t) &=& \exp\left[ - \frac{\pi}{3}  \epsilon v^2  t^3 \right],  
{\rm ~for~}  t < t^*,\nonumber \\
& = & \exp\left[ - \pi \epsilon v^2 {t^*}^2 \left( t- \frac{2 t^*}{3 } 
\right) \right], {\rm ~for~} t > t^*.
\label{eq:nucleation}
\end{eqnarray}

We see that with this choice, both $Q(t)$ and its derivative are 
continuous at $t = t^*$. Since $V(t')$ should tend to $L^2$ for large 
$t'$, we get the crossover scale $t^*$ given by
\begin{equation}
t^* = \frac{L}{v \sqrt{\pi}}.
\end{equation}
The crossover lattice size $L^*$ beyond which the average lifetime of 
the metastable state becomes independent of $L$ can then be estimated 
from the above to be
\be 
L^*  \sim \left(\frac{3 \sqrt{\pi} v}{ \epsilon} \right)^{1/3}.
\label{eq:critL}
\ee

Fitting the numerical data in Fig.~\ref{fig:Q_tK7L154} to 
Eq.~(\ref{eq:nucleation}) we obtain $\epsilon= (2.1 \pm 0.2) \times 
10^{-10}$ and $v=(5.5 \pm 0.7) \times 10^{-5}$ for $\mu=7.60$. From 
Eq.~(\ref{eq:critL}), we then obtain the crossover scale $L^* \sim 110$, 
of the same order as the numerically observed value of $L^*\sim 200$.  
The difference is presumably due to simplifying approximations  made in the 
theory, 
e.g., neglecting the dependence of the mean velocity of growth on the 
direction of growth, or the curvature of the interface, etc.

We can also estimate $v$ directly from simulations of a system with an 
initial configuration where half the sample is in the nematic phase and 
the other half is in the equilibrium disordered phase at that $\mu$. For 
$\mu=7.60$, we find that this velocity increases slowly with $L$, and tends to 
a limiting value  $\approx 1.0 \times 10^{-4}$  for 
$L \geq 784$, reasonably close to the velocity obtained from fitting data to 
Eq.~(\ref{eq:nucleation}).  For decreasing chemical potential $\mu$, we 
find that both the velocity $v$ and nucleation rate $\epsilon$ increase.

\section{\label{sec:highdensity} \bf Nature of the high-density
disordered phase}

There is a one-to-one 
correspondence  between fully packed k-mer configurations and a restricted 
solid on solid height model with vector-valued
heights~\cite{kenyon1992,thurston1990}. The height fluctuations 
at large length scales are well-described by a gaussian model, and
at full packing the orientation-orientation correlation function 
decays as  a power-law with distance. The exponent of this power law has 
been  estimated for the case $k=3$ by exact diagonalization 
studies~\cite{ghosh2}.
If these correlations are not destroyed by small density of 
vacancies for large $k$, then the 
correlations in the HDD phase  would 
be  long-ranged, qualitatively different from the known exponential decay of 
correlations in the LDD phase. In this section, we test 
this possibility by studying the susceptibility $\chi$, the 
order parameter correlation function $C_{SS}(i,j)$, the  cluster size 
distribution $F(s)$, and the size distribution of 
structures that we call stacks. We also examine the formation of bound 
states of vacancies.

\begin{figure}
\includegraphics[width=\columnwidth]{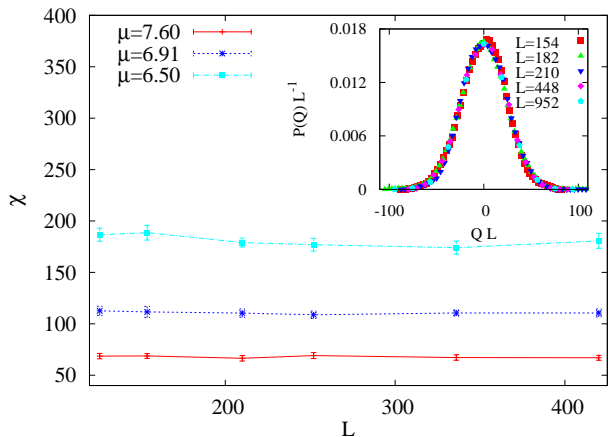}
\caption{(Color online) Susceptibility $\chi$  for the square lattice
as a function of $L$ for three values
of $\mu$, all in the HDD phase.
There is no anomalous dependence on $L$. Inset: The scaled probability
distribution for the order parameter $P(Q)$ for different $L$'s collapse when
plotted against $Q L$.  The data are for $\mu=5.95$.}
\label{fig:chilargemu}
\end{figure}

The susceptibility is defined as $\chi = L^2 \langle (n_h-n_v)^2 
\rangle/\langle n_h+n_v \rangle^2$, where $n_h$ and $n_v$ are the number 
of x-mers and y-mers. Figure~\ref{fig:chilargemu} shows the variation of 
$\chi$ with $L$, for three different values of $\mu$ in the HDD phase. 
$\chi$ tends to a finite non-zero value for large $L$, hence, 
if the correlations are 
a power law, then the decay exponent is larger than 2.  From the
central limit theorem, it follows that 
the order parameter $Q$ should scale as $L^{-1}$.  This is 
confirmed in 
the inset of Fig.~\ref{fig:chilargemu}, where the scaled probability 
distributions for different $L$'s collapse onto one curve when plotted 
against $Q L$.

\begin{figure}
\includegraphics[width=\columnwidth]{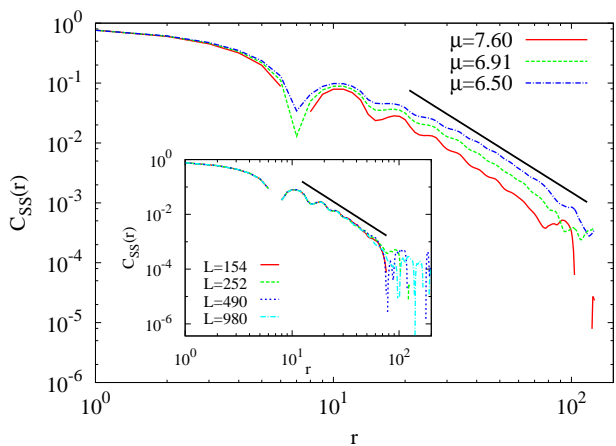}
\caption{(Color online) Order parameter correlations $C_{SS}(r)$  for the 
square lattice as a function of $r$, measured along the $x$- and $y$-axes, 
for three 
different values of $\mu$, all of them larger than $\mu_c\approx 
5.57$. The system size is $L=252$. 
Inset: The dependence of $C_{SS}(r)$ on $L$ is shown for 
$\mu=7.60$. The solid lines are power laws $r^{-2.5}$, intended only as 
guides to the eye.} \label{fig:fig_corr_order_mu}
\end{figure}

The order parameter correlation function  $C_{SS}(i,j)$ 
is defined as 
follows. Given a configuration, we assign to each site $(i,j)$ a 
variable $S_{i,j}$, where $S_{i,j}=1$ if $(i,j)$ is occupied by an 
x-mer, $S_{i,j}=-1$ if $(i,j)$ is occupied by an y-mer, and $S_{i,j}=0$ 
if $(i,j)$ is empty. Then,
\be
C_{SS}(i,j)= \langle S_{0,0} S_{i,j} \rangle.
\ee
Figure~\ref{fig:fig_corr_order_mu} shows the variation of $C_{SS}(r)$ 
with separation $r$ along the $x$- and $y$- axes, for different chemical 
potentials and systems sizes. In the HDD phase, the 
correlation function has an oscillatory dependence  on  distance with 
period $k$, and for $r \gg k$, appears to decrease as a power law 
$r^{-\eta}$, with $\eta>2$. Given the limited range of $r$ available $7 
\ll r \ll L/2$, it is difficult to get an accurate estimate of the 
exponent $\eta$.

The long-range correlations in the HDD phase are better studied by 
looking at the large-scale properties of connected clusters of parallel 
rods. For instance, it is known that the exponent characterizing the 
decay of cluster size distribution of critical Fortuin--Kasteleyn 
clusters \cite{FK1972} in the $q$-state Potts model 
\cite{Janke2004,Deng2004} has a non-trivial dependence on $q$. We denote 
all sites occupied by x-mers by $1$ and the rest by zero. For our 
problem, we define a cluster as a set of $1$'s connected by nearest 
neighbor bonds.  Let $F(s)$ be the probability that a randomly chosen 
$1$ belongs to a cluster of $s$ sites.  Clearly, $F(s)$ is zero, unless 
$s$ is a multiple of $k$. Let the cumulative distribution function be 
$F_{cum}(s) = \sum_{s'=1}^s F(s')$.

\begin{figure} 
\includegraphics[width=\columnwidth]{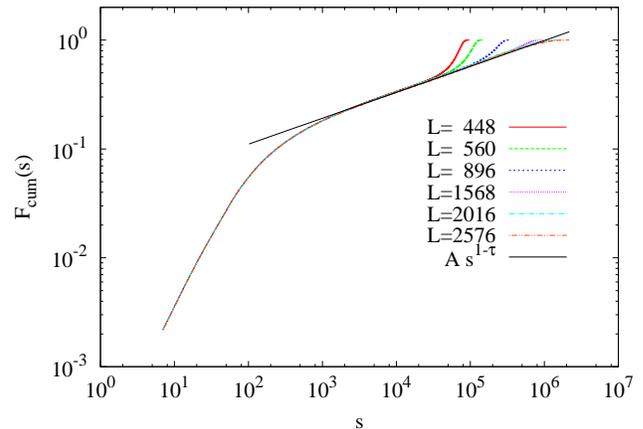} 
\caption{(Color online) $F_{cum}(s)$, the probability that a randomly chosen 
$1$ (a site occupied by a x-mer) belongs to a connected cluster of size
$\leq s$, in the HDD phase ($\mu=7.60$) for different system sizes.
The data are for the square lattice.
}
\label{fig:clusterdist} 
\end{figure}

In Fig.~\ref{fig:clusterdist}, we plot $F_{cum}(s)$ in the HDD phase for 
different system sizes on the square lattice.  We find that for 
intermediate range of $s$, for 
$10^3 \ll s \ll 10^6$, $F_{cum}(s) \simeq A s^{1-\tau}$, with $\tau<1$. 
For $\mu =7.60$, we estimate the numerical values to be $A=0.037$ and 
$\tau=0.762$. For small system sizes (up to $L=1568$), $F_{cum}(s)$ has 
a system-size dependent cutoff. The $L$-independent cutoff $s^*$ is 
determined by the condition $A s^{* 1-\tau} \approx 1$, giving $s^* 
\approx 1.04\times 10^6$. The density of $1$'s being roughly 0.48, we 
expect to observe $s^*$ only when $L$ exceeds a characteristic length 
scale $\xi^* \sim 1400$. This is indeed seen in 
Fig.~\ref{fig:clusterdist}.

\begin{figure} 
\includegraphics[width=\columnwidth]{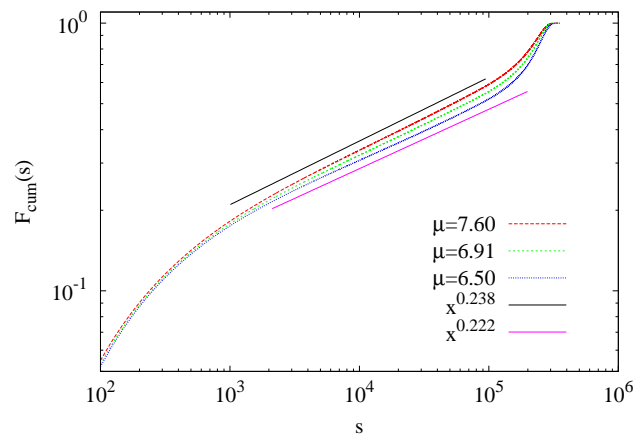} 
\caption{(Color online) $F_{cum}(s)$, the probability that a randomly 
chosen $1$ (a site occupied by a x-mer) belongs to a connected cluster 
of size $\leq s$, for different values of $\mu$, all corresponding to the HDD 
phase.  The curves appear to have weakly density dependent power-law 
exponents.}
\label{fig:clusterdist_mu} 
\end{figure}

In the HDD phase, $F_{cum}(s)$ depends weakly on $\mu$ (see 
Fig.~\ref{fig:clusterdist_mu}). The power law exponent $\tau$ is 
estimated to be $0.778$ ($\mu=6.50$), $0.767$ ($\mu=6.91$) and $0.762$ 
($\mu=7.60$). It appears that $\tau$ decreases  slowly with increasing 
$\mu$, while $s^*$  decreases with increasing $\mu$.

\begin{figure}
\begin{center}
\begin{tabular}{cc}
\includegraphics[width=4.15cm]{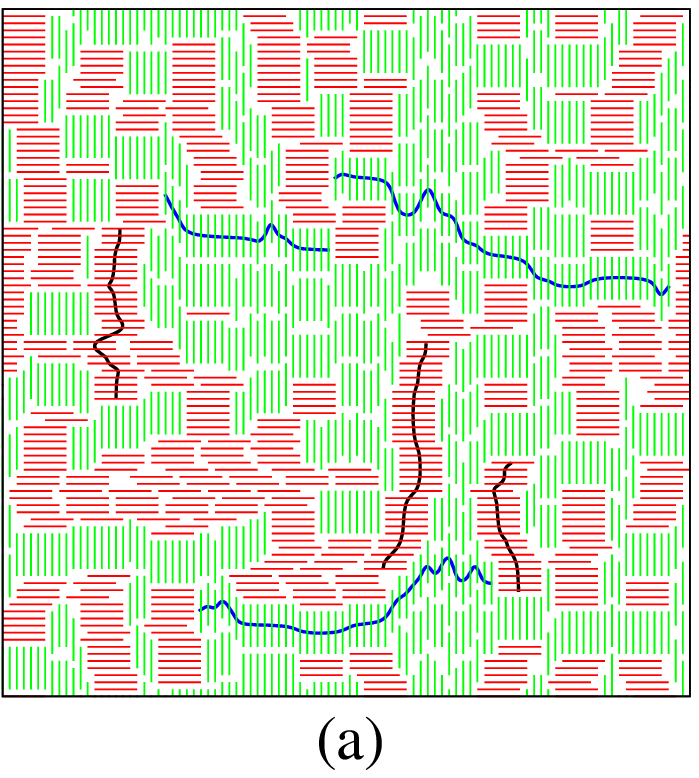}
\includegraphics[width=4.15cm]{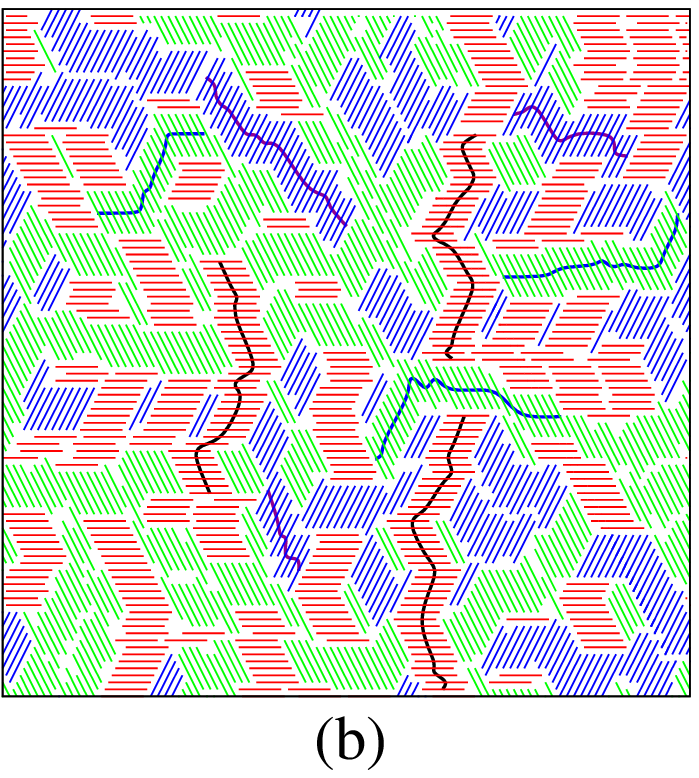}
\end{tabular}
\caption{(Color online) Some examples of the different types of
stacks, shown here as rods joined by wiggly lines,   
for (a) square lattice and (b) triangular lattice.
The snapshots are for $\mu=7.60$, corresponding to the 
HDD phase. Rods of different orientations are shown in different colors
for easy visualization.
}
\label{fig:fig_stack}
\end{center}
\end{figure}

One qualitative feature of the HDD phase is the appearance of large 
groups of parallel rods, worm-like in appearance, nearly aligned in the 
transverse direction. This is clearly seen in 
Fig.~\ref{fig:fig_snapshot}(c). We call these groups stacks. To be 
precise, we define a stack as follows: two neighboring parallel $k$-mers 
are said to belong to the same stack if the number of nearest-neighbor 
bonds between them is greater than $k/2$.  A stack is the maximal 
cluster of rods that can be so constructed.  By this definition, a stack 
has a linear structure without branching, with some transverse 
fluctuations allowed. Examples of stacks on square and triangular 
lattices are shown in Fig.~\ref{fig:fig_stack}.  Any given configuration 
of rods is uniquely broken up into a collection of disjoint stacks.

\begin{figure}
\includegraphics[width=\columnwidth]{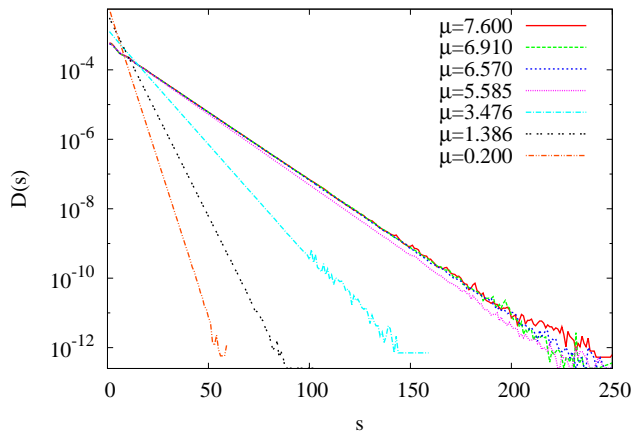}
\caption{(Color online) Stack distribution in the LDD phase 
($\mu = 0.200$), intermediate density nematic phase ($\mu = 3.476$), 
HDD phase ($\mu = 7.600$), and at 
two critical points ($\mu = 1.3863$, $5.57$) are shown. 
Data are for $L=280$, $k=7$, and the square lattice.}
\label{fig:stackdist}
\end{figure}

There are a noticeable number of large stacks in the HDD phase. We 
measured the stack size distribution $D(s)$, the number of stacks of 
size $s$ per site of the lattice, in all the three phases and at the 
transition points (see Fig.~\ref{fig:stackdist}).  Interestingly, we 
found that this distribution is nearly exponential in all the three 
phases, as well as at the critical points, and there is no indication of 
any power-law tail in this function.  In the HDD phase, the mean stack 
size is approximately $12$, for both square and triangular lattices, and 
is only weakly dependent on the density.

It was suggested in Ref.~\cite{ghosh2007} that the second phase 
transition may be viewed as a binding-unbinding transition of $k$ 
species of vacancies.  For studying such a characterization, we break 
the square lattice into $k$ sublattices. A site $(x,y)$ belongs to the 
$i$-th sublattice if $x+y = i~(mod~k)$, where $i = 0, 1, \ldots, k-1$. 
In a typical configuration with a low density of vacant sites, it was 
argued that the vacancies would form bound states of $k$ vacancies, one 
from each sublattice. The HDD phase can then be described as a weakly 
interacting gas of such bound states if the typical distance between two 
bound states is much larger than the mean size of a bound state.

Let $d_{ij}$ be the Euclidean distance between a randomly picked vacant 
site on the $i$-th sublattice, and the vacant site nearest to it on the 
$j$-th sublattice. The average of $d_{ij}$, averaged over all pairs 
$(ij)$, with $i \neq j$, will be denoted by $\bar{d}_{ij}$, and 
$\bar{d}_{ii}$ will denote the value of $d_{ii}$, averaged over $i$.

\begin{figure}
\includegraphics[width=\columnwidth]{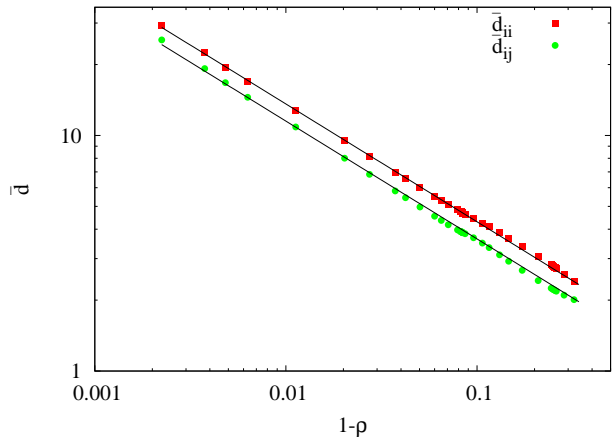}
\caption{(Color online) The average  spacing between vacancies
$\bar{d}_{ii}$ and $\bar{d}_{ij}$, on the square lattice as a function of 
density $\rho$. The solid lines show the functions $K (1-\rho)^{-1/2}$, 
for $K=1.36$ and $1.15$. The data are for $L=168$ and $k=7$.}
\label{fig_unbinding}
\end{figure}

In Fig.~\ref{fig_unbinding}, we show the variation of $\bar{d}_{ij}$ and 
$\bar{d}_{ii}$ with density $\rho$.  We see that $\bar{d}_{ii}$ and 
$\bar{d}_{ij}$, both vary approximately as $(1 - \rho)^{-1/2}$, with 
$\bar{d}_{ii} \approx 1.18 \bar{d}_{ij}$. The data are for $L=168$ and 
$k=7$. There is no noticeable dependence of the data on $L$. We see no 
signature of $\bar{d}_{ij}$ saturating to a finite value, for the 
densities up to $0.995$, when $\bar{d}_{ij} \simeq 35$.

We conclude that  the bound state, if exists at all, is very weakly bound.   
Near $\rho_2^*$, the typical spacing between vacancies is much less than  
the size of the bound state, and  the transition  can not be treated as 
binding-unbinding transition when the average distance between bound states  
becomes comparable to their size.

\section{\label{sec:results}Critical behavior near the second transition}

We now discuss the critical behavior near the second transition. Several 
thermodynamic quantities are of interest. We define the nematic order 
parameter $m$ as follows. For the square lattice, $m= (n_h -n_v)/N$, 
where $n_h$ and $n_v$ are the number of lattice sites occupied by x-mers 
and y-mers respectively, and $N$ is the total number of lattice sites. 
For the triangular lattice, $m = (n_1 + \omega n_2 + \omega^2 n_3)/N$, 
where $\omega$ is the complex cube-root of unity, and $n_1, n_2, n_3$ 
are the number of sites occupied by $k$-mers oriented along the three 
allowed directions. The density $\rho$ is defined by the fraction of 
sites that are occupied by the $k$-mers. The order parameter $Q$, its 
second moment $\chi$, compressibility $\kappa$, and Binder cumulant $U$ 
are defined as
\begin{subequations}
\label{eq:thermodynamics}
\bea
Q&=&\frac{\langle |m| \rangle}{\langle \rho \rangle},\\
\chi & =& \frac{L^2 \langle |m|^2 \rangle}
{\langle \rho\rangle^2},\\
\kappa & =& L^2 \left[\langle \rho^2 \rangle -\langle \rho \rangle^2
\right],\\
U & =& 1- \frac{ \langle |m|^4 \rangle} { a \langle |m|^2 \rangle^2},
\eea
\end{subequations} 
where $a=3$ for square lattice and $a=2$ for triangular lattice. 
$Q$ is zero in the LDD and HDD phases and nonzero in the 
nematic phase.

\begin{figure} 
\includegraphics[width=\columnwidth]{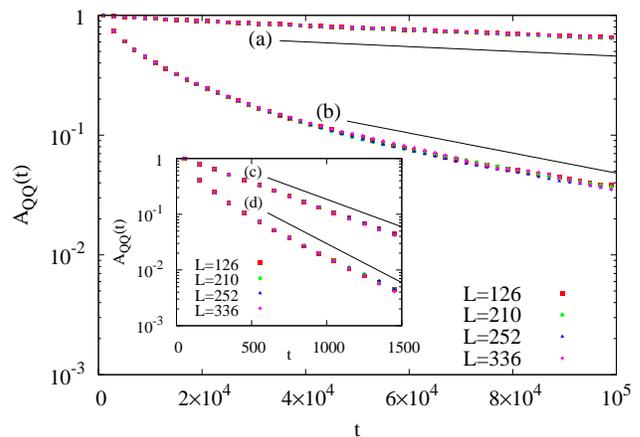} 
\caption{(Color online) The temporal variation of the autocorrelation 
functions for (a) the global order parameter $Q$ and (b) the local order 
parameter $S$. The data are for $\mu=7.60$ and the autocorrelation times 
corresponding to the solid lines are (a)220000 and (b) 52000. Inset: 
Data as above but in the LDD phase ($\mu=0.20$), with (c) corresponding 
to $A_{QQ}(t)$ (d) corresponding to $A_{SS}(t)$. The autocorrelation 
times are (c) 440 and (d) 310. All data are for $k=7$. }
\label{fig:autocorrelation} 
\end{figure}

The data used for estimating the critical exponents are for $k=7$, and 
for five different system sizes $L=154$, $210$, $336$, $448$, and $952$ 
for the square lattice and $L=210$, $336$, $448$, and $560$ for the 
triangular lattice. The system is equilibrated for $10^7$ Monte Carlo 
steps for each $\mu$, following which the data are averaged over $3 
\times 10^8$ Monte Carlo steps. These times are larger than the largest 
autocorrelation times that we encounter obtained by
measuring the autocorrelation function
\be
A_{QQ}(t) = \frac{\langle Q(\tau) Q(\tau+t) \rangle}{\langle Q^2
\rangle},
\label{eq:autocorrelation}
\ee
where the averaging is done over the reference time $\tau$. 
The function $A_{SS}(t)$ is defined similarly.
The largest autocorrelation time is for 
the largest density and is close to $2.2 \times 10^5$ Monte Carlo
steps (see Fig.~\ref{fig:autocorrelation}). 
To estimate errors, the measurement is broken up into 10 statistically 
independent blocks. 

The quantities in Eq.~(\ref{eq:thermodynamics}) are determined as a 
function of $\mu$ using Monte Carlo simulations. The nature of the 
second transition from the ordered nematic phase to the HDD phase 
is determined by the singular behavior of $U$, $Q$, $\chi$, and $\kappa$ 
near the critical point. Let $\epsilon = (\mu-\mu_c)/\mu_c$, where 
$\mu_c$ is the critical chemical potential. The singular behavior is 
characterized by the critical exponents $\nu,$ $\beta$, $\gamma$, and 
$\alpha$, defined by $Q \sim (-\epsilon)^\beta$, $\epsilon<0$, $\chi \sim
|\epsilon|^{-\gamma}$ and $\kappa \sim
|\epsilon|^{-\alpha}$, and $\xi^* \sim |\epsilon|^{-\nu}$, 
where $\xi^*$ is the correlation length and $|\epsilon| \rightarrow 0$. The 
exponents are obtained by finite size scaling of the different quantities 
near the critical point:
\begin{subequations}
\label{eq:scaling}
\bea
U &\simeq&  f_U(\epsilon L^{1/\nu}), \label{eq:Uscaling}\\
Q &\simeq& L^{-\beta/\nu} f_Q(\epsilon L^{1/\nu}), \label{eq:Qscaling}\\
\chi & \simeq& L^{\gamma/\nu} f_{\chi}(\epsilon L^{1/\nu}),
\label{eq:chiscaling}\\
\kappa & \simeq & L^{\alpha/\nu} f_\kappa(\epsilon L^{1/\nu}),
\label{eq:kappascaling}
\eea
\end{subequations}
where $f_U$, $f_Q$, $f_{\chi}$, and $f_\kappa$ are scaling functions.

\subsection{Square lattice}

\begin{figure}
\includegraphics[width=\columnwidth]{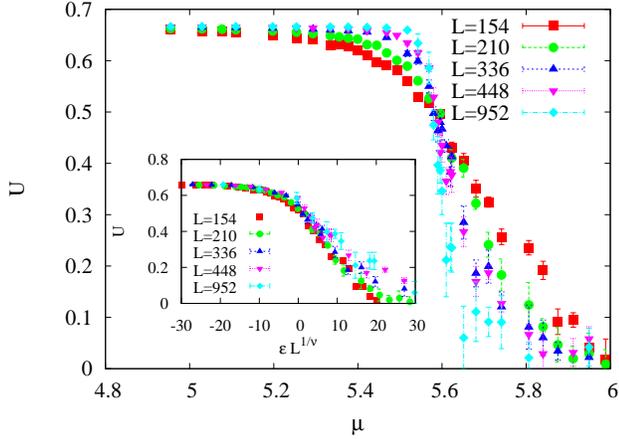}
\caption{(Color online) The Binder cumulant $U$ as a function of chemical potential $\mu$ 
for different lattice sizes of a square lattice. The curves intersect at 
$\mu_c = 5.57 \pm .02$.
Inset: Data collapse for square lattices when $U$ is plotted against 
$\epsilon L^{1/\nu}$ with $\nu=0.90$ and $\epsilon = (\mu- 
\mu_c)/\mu_c$.
} 
\label{fig:fig_second_u}
\end{figure}

We first present results for the square lattice. The data for the Binder 
cumulant $U$ for different system sizes intersect at $\mu_c 
= 5.57 \pm .02$  (see Fig.~\ref{fig:fig_second_u}). The density at this 
value of chemical potential is $\rho_2^* = 0.917 \pm .015$, consistent 
with the variational estimate $ 0.87 \leq \rho_2^* \leq 0.93$ in 
Ref.~\cite{linares2008}. The value of $U$ at the transition lies in the 
range $0.56$ to $0.59$. This is not very different from the value for 
the Ising transition $U_c \approx 0.61$ \cite{blote1993}. The data for 
different system sizes collapse when scaled as in 
Eq.~(\ref{eq:Uscaling}) with $\nu = 0.90 \pm .05$ (see inset of 
Fig.~\ref{fig:fig_second_u}). To compare with the first transition from 
the LDD phase to the nematic phase, $\rho_1^*= 0.745 \pm 0.010$, and the 
numerical estimate for the exponent $\nu$ is consistent 
with the known exact Ising value $1$ \cite{fernandez2008a}.

\begin{figure}
\includegraphics[width=\columnwidth]{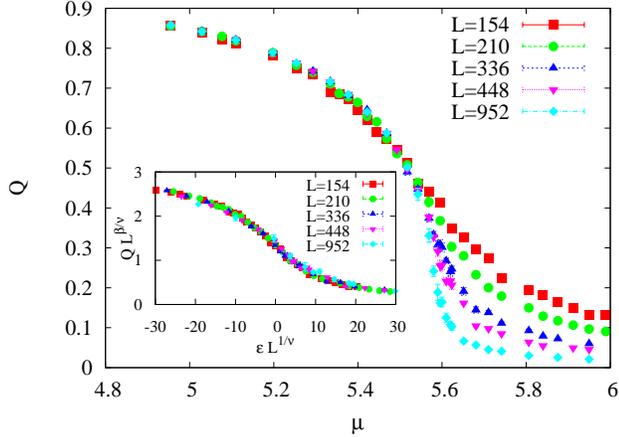}
\caption{(Color online) The variation of the order parameter $Q$ with chemical 
potential $\mu$ for different systems sizes of a square lattice. Inset: 
Data collapse for square lattices when scaled $Q$ is plotted against 
$\epsilon L^{1/\nu}$ with $\nu=0.90$, $\beta/\nu =0.22$ and $\epsilon = 
(\mu- \mu_c)/\mu_c$. 
} 
\label{fig:fig_second_order}
\end{figure}

\begin{figure}
\includegraphics[width=\columnwidth]{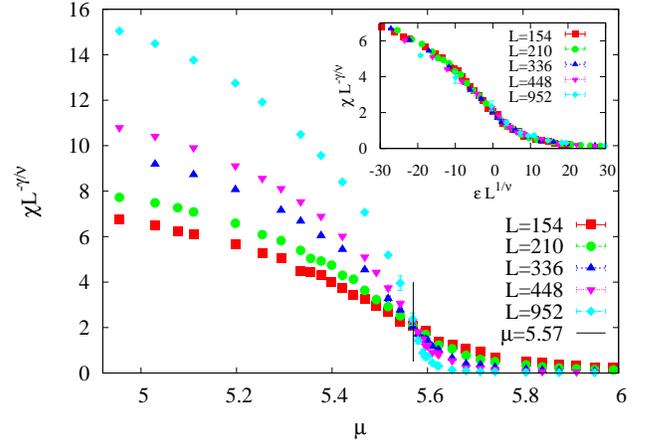}
\caption{(Color online) The variation of $\chi$, the mean of the square of 
the order 
parameter, with chemical potential $\mu$ for different system sizes of a 
square lattice. The curves cross at $\mu_c$ when $\chi$ is scaled by 
$L^{-\gamma/\nu}$, with $\gamma/\nu=1.56$. Inset: Data collapse for 
square lattices when $\chi L^{-\gamma/\nu}$ is plotted against $\epsilon 
L^{1/\nu}$ with $\nu=0.90$, and $\epsilon = (\mu- \mu_c)/\mu_c$. 
} 
\label{fig:fig_second_chi}
\end{figure}

\begin{figure}
\includegraphics[width=\columnwidth]{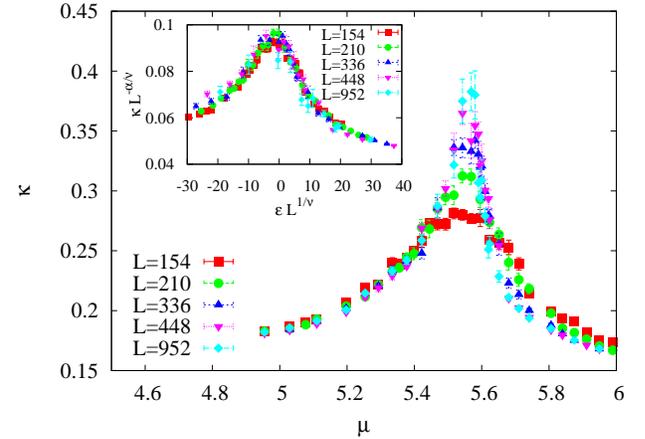}
\caption{(Color online) The variation of compressibility $\kappa$ with chemical 
potential $\mu$ for different system sizes of a square lattice. Inset: 
Data collapse for square lattices when the scaled $\kappa$ is plotted 
against $\epsilon L^{1/\nu}$ with $\nu=0.90$, $\alpha/\nu=0.22$, and 
$\epsilon = (\mu-\mu_c)/\mu_c$.
} 
\label{fig:fig_second_comp}
\end{figure}

The data for order parameter, $\chi$ and $\kappa$ 
for different system sizes are shown in 
Figs.~\ref{fig:fig_second_order}, \ref{fig:fig_second_chi} and
\ref{fig:fig_second_comp} respectively. $Q$ decreases to zero at high 
densities. 
Our best estimates of effective critical exponents are
$\beta/ \nu = 0.22 \pm 0.07$ (see inset of Fig.~\ref{fig:fig_second_order}).
$\gamma /\nu = 1.56 \pm 0.07$ (see inset of Fig.~\ref{fig:fig_second_chi}), 
and $\alpha/ \nu = 0.22 \pm 0.07$ (see inset of
Fig.~\ref{fig:fig_second_comp}).
The estimated error bars are our subjective estimates, 
based on the goodness of fit. These differ substantially from the
values of the exponents of the two dimensional Ising model 
($\nu = 1$, $\beta= 1/8, \gamma = 7/4, \alpha = 0$).
However, as discussed  in Sec.~\ref{sec:highdensity},
it seems like there is a characteristic length scale $ \xi^*
\sim 1400$ in the HDD phase, and we cannot say much about the 
asymptotic value the critical exponents at length scales 
$L \gg \xi^*$.

\subsection{Triangular lattice}

\begin{figure}
\begin{tabular}{cc}
\includegraphics[width=4.15cm]{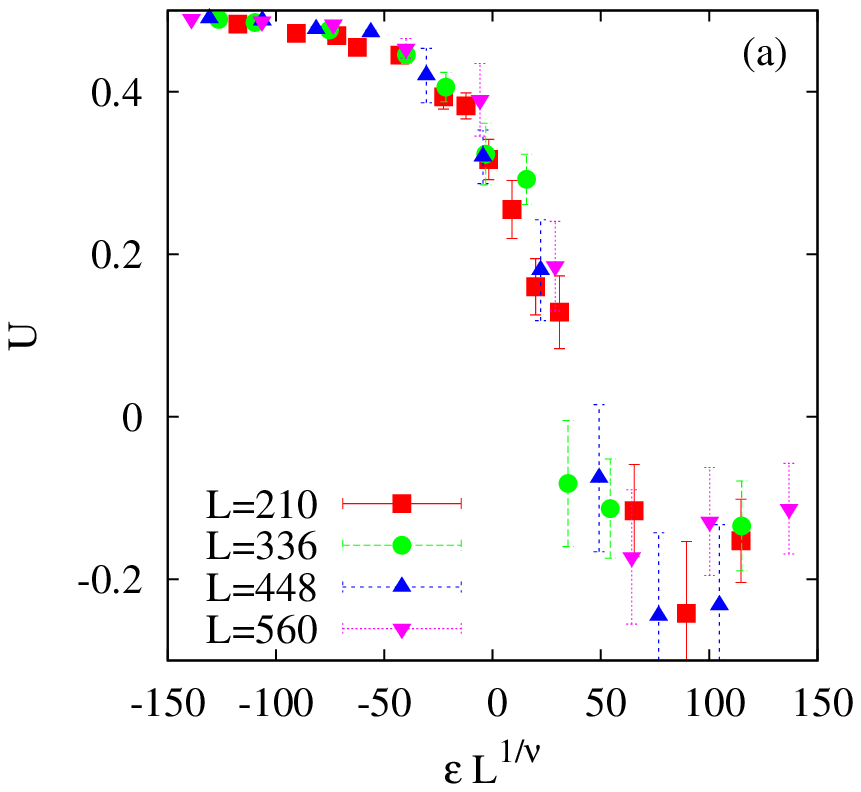}
\includegraphics[width=4.15cm]{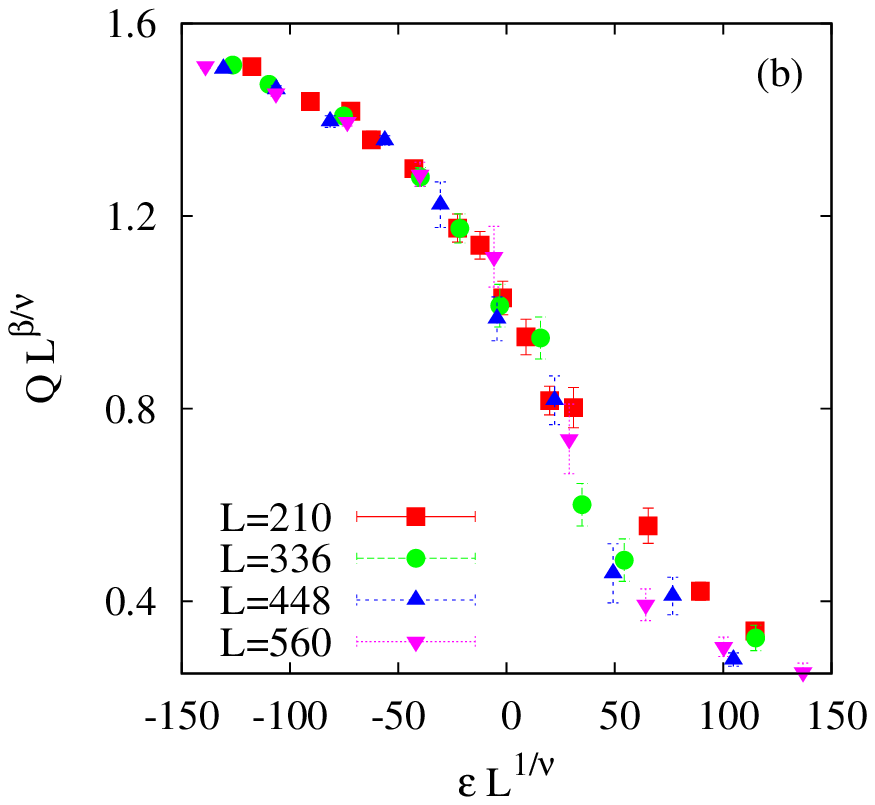}\\
\includegraphics[width=4.15cm]{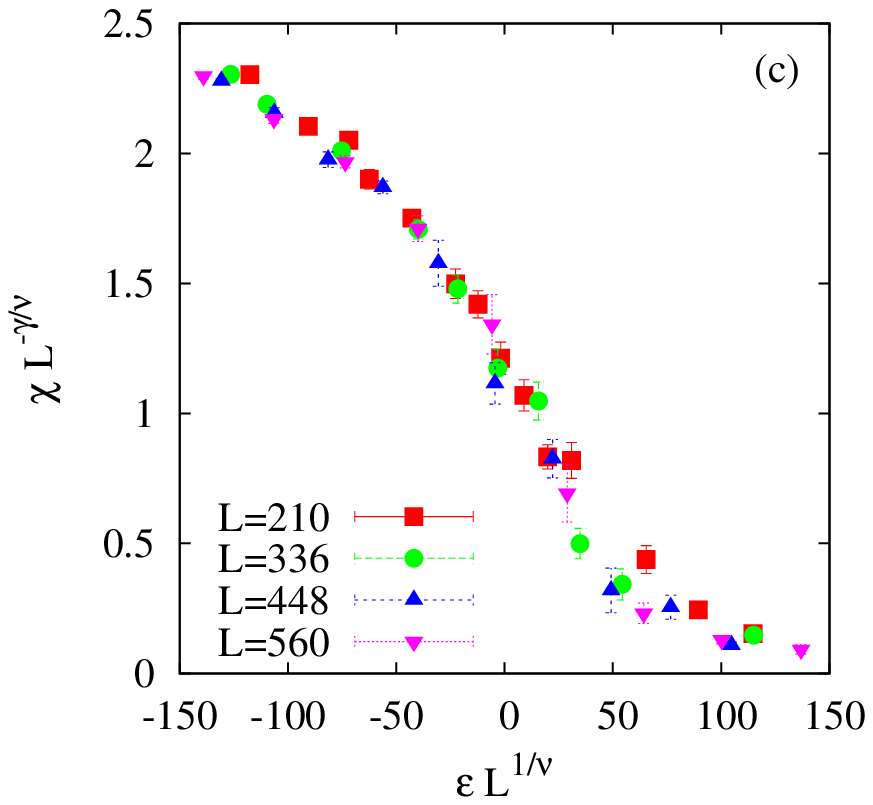}
\includegraphics[width=4.15cm]{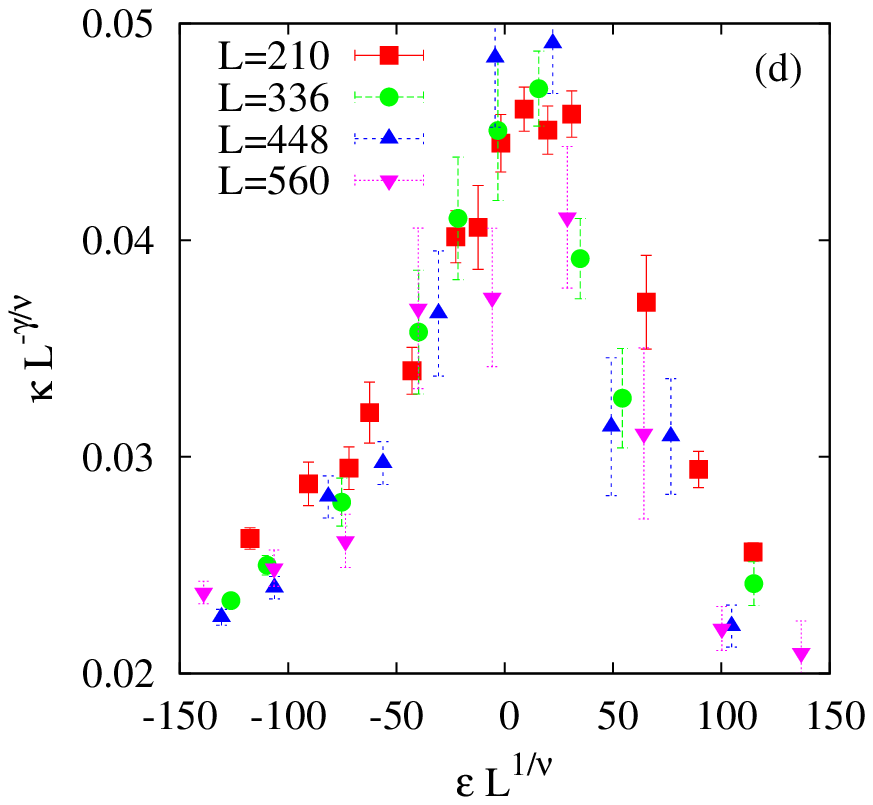}
\end{tabular}
\caption{(Color online) Finite size scaling of the triangular lattice data of
(a) U, (b) Q, (c) $\chi$, and (d) $\kappa$ 
as in Eq.~(\ref{eq:scaling}) with $\nu=5/6$, $\beta/\nu=2/15$, 
$\gamma/\nu=26/15$ and $\alpha/\nu=2/5$. The critical chemical potential is
$\mu_c=5.147$.}
\label{fig:fig_tri_scaling}
\end{figure}

For the triangular lattice,
we find that the second transition is 
continuous with $\mu_c = 5.147\pm .050$ and $\rho^*_2 = .905 \pm .010$. 
The data for $U$, $Q$, $\chi$, and $\kappa$ for different system sizes 
collapse onto one scaling curve when scaled as in Eq.~(\ref{eq:scaling}) 
with exponents that are indistinguishable from those of the three state 
Potts model (see Fig.~\ref{fig:fig_tri_scaling}) ($\nu=5/6$, $\beta=1/9$,
$\gamma =13/9$ and $\alpha=1/3$). 

\begin{figure} 
\includegraphics[width=\columnwidth]{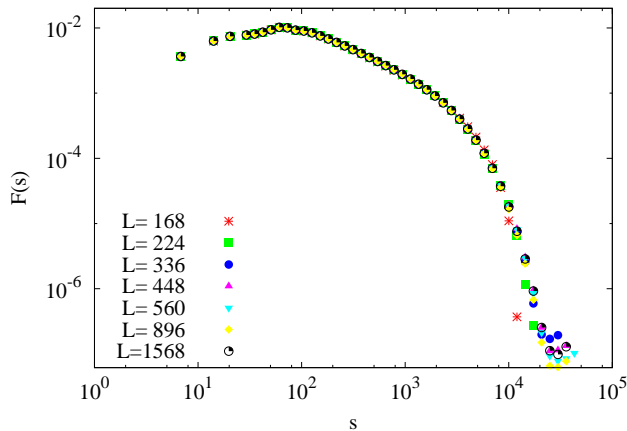} 
\caption{(Color online) $F(s)$, the probability 
that a randomly chosen site belongs to a $s$-site connected cluster of 
horizontal rods, in the HDD phase for different system sizes. The data
are for the triangular lattice for $\mu=7.60$.}
\label{fig:tri_cluster} 
\end{figure}

As in the case of the square lattice, we probe the 
correlations in the triangular lattice by looking at the large-scale 
properties of connected clusters of parallel rods. We denote all sites 
occupied by horizontal rods by $1$ and the rest by zero. In 
Fig.~\ref{fig:tri_cluster}, $F(s)$, the probability that a randomly 
chosen $1$ belongs to a cluster of $s$ sites, is shown for different 
system sizes $L$ in the HDD phase. Unlike the square 
lattice case, here there is no extended regime of $s$ where $F(s)$ seems 
to grow as a power of $s$. This suggests that for the triangular 
lattice, the HDD and LDD phases are qualitatively similar, and the HDD 
phase has a finite correlation length $\sim 60$.

\section{\label{sec:summary}Summary and Discussion }

In this paper, we studied the problem of hard, rigid rods on two 
dimensional square and triangular lattices, using an efficient algorithm 
that is able to overcome jamming at high densities.  The algorithm is 
more efficient than algorithms with only local moves. In addition to 
overcoming jamming at high packing densities, it is easily parallelized, 
which makes it suitable for studying hard-core systems with other 
particle shapes, and also in higher dimensions.

We showed the existence of a second transition from the ordered nematic 
phase to a disordered phase as the packing density is increased. By 
studying the order parameter, its second moment, compressibility and 
Binder cumulant, we find that the second transition is continuous on 
both square and triangular lattices. We also investigated the nature of 
correlations in the HDD phase by measuring distribution of connected 
clusters of parallel rods, as well as the distribution of stacks.

We are not able to give a very clear answer to the question 
whether the HDD and LDD phases are qualitatively different, or not.  But 
the available evidence suggests that while the HDD phase has a large 
correlation length $\xi^*$, it is not qualitatively different. This is 
based on the evidence that vacancies in the HDD phase do not form a 
bound state. In that sense, a k-mer system with $k=7$, at high densities 
is similar to the $k=2$ case, where also, the monomers do not form a 
bound state, and unbound monomers lead to an exponential decay of 
correlations at any non-zero vacancy density.

Additional support for this scenario comes from the fact that if
the hard-core constraint is relaxed, and  two k-mers are
allowed to share a site, but with a cost in energy, then exact calculation 
on an artificial lattice (the random locally tree-like-layered lattice) shows 
\cite{kundu2013} two phase transitions at densities $\rho_{c1}$ and 
$\rho_{c2}$ for a range of values of the repulsive energy.
The difference between these critical densities
decreases as the repulsive energy is decreased, 
and below a particular value of the repulsive energy, the intermediate 
nematic phase disappears. Thus, in this special case, which is the only 
known exactly solved model of $k$-mers showing two phase transitions, 
the LDD and HDD phases are the same.

An interesting feature of the HDD phase is the appearance of a large 
characteristic length scale $\xi^* \sim 1400$ on the square lattice, as 
inferred from the fact that the cluster size distribution seems to 
follow a power law distribution $F(s) \sim s^{-\tau}$, with $\tau < 1$ 
for $s < \xi^{*2}$. The amplitude of this power-law term is rather small. 
This is related to the fact that for the $k$-mer problem, various 
perturbation series involve terms like $k^{-k}$ \cite{ghosh2007}, which 
then leads to large correlation lengths. The HDD phase has power-law 
correlations at least for lengths up to $\xi^*$.

For the triangular lattice, the correlation length $\xi^*$ 
is much smaller, as near the critical point, clusters of each type of 
rods cover only about a third of the sites, which is substantially below 
the corresponding percolation threshold.

On the square lattice, our best estimates of the numerical values of the 
critical exponents are different from those of the Ising universality 
class. However, because the correlation lengths in the HDD phase are 
large, we cannot rule out a crossover to the Ising universality class at 
larger length scales. For the triangular lattice with $k=7$, the 
estimated exponents for second transition are consistent with those of 
two dimensional 3-state Potts universality class.

We expect that the nature of the transition will not depend on the rod 
length $k$.  For example, the first transition at $\rho_{1}^*$ is 
Ising-like, whenever it occurs. Several interesting questions remain. 
For example, the entropy per site in the fully packed $k$-mer problem is 
expected to vary as $( c \log k)/ k^2$, for $k$ large, where $c$ is a 
constant. The argument in Ref.~\cite{ghosh2007} relates this constant to 
the coefficient $a$ that appears to the way $\rho_2^*$ varies with $k$, 
for large $k$: $\rho_2^* = 1 - a/k^2 + \ldots$ for large $k$.  It would 
be interesting to determine $c$ and $a$ exactly, or to test these by 
simulations with larger $k$.  Also, further studies are needed to 
determine if the correlations in the HDD phase decay exponentially for 
distances greater than $\xi^*$. This seems like a promising area for 
further study.

\begin{acknowledgments}
We would like to thank K. Damle, P. Grassberger, S. Digal, 
D. Das, P. Ray and G. I. Menon for helpful discussions.
The simulations for the larger lattice sizes were carried out on the
supercomputing
machine Annapurna (Intel Nehalem 2.93 GHz) at The Institute of 
Mathematical Sciences.
\end{acknowledgments}


\begin{thebibliography}{31}
\expandafter\ifx\csname natexlab\endcsname\relax\def\natexlab#1{#1}\fi
\expandafter\ifx\csname bibnamefont\endcsname\relax
  \def\bibnamefont#1{#1}\fi
\expandafter\ifx\csname bibfnamefont\endcsname\relax
  \def\bibfnamefont#1{#1}\fi
\expandafter\ifx\csname citenamefont\endcsname\relax
  \def\citenamefont#1{#1}\fi
\expandafter\ifx\csname url\endcsname\relax
  \def\url#1{\texttt{#1}}\fi
\expandafter\ifx\csname urlprefix\endcsname\relax\def\urlprefix{URL }\fi
\providecommand{\bibinfo}[2]{#2}
\providecommand{\eprint}[2][]{\url{#2}}

\bibitem[{\citenamefont{Onsager}(1949)}]{onsager1949}
\bibinfo{author}{\bibfnamefont{L.}~\bibnamefont{Onsager}},
  \bibinfo{journal}{Ann. N.Y. Acad. Sci.} \textbf{\bibinfo{volume}{51}},
  \bibinfo{pages}{627} (\bibinfo{year}{1949}).

\bibitem[{\citenamefont{Flory}(1956)}]{flory1956a}
\bibinfo{author}{\bibfnamefont{P.~J.} \bibnamefont{Flory}},
  \bibinfo{journal}{Proc. R. Soc.} \textbf{\bibinfo{volume}{234}},
  \bibinfo{pages}{60} (\bibinfo{year}{1956}).

\bibitem[{\citenamefont{Zwanzig}(1963)}]{zwanzig1963}
\bibinfo{author}{\bibfnamefont{R.}~\bibnamefont{Zwanzig}}, \bibinfo{journal}{J.
  Chem. Phys.} \textbf{\bibinfo{volume}{39}}, \bibinfo{pages}{1714}
  (\bibinfo{year}{1963}).

\bibitem[{\citenamefont{Vroege and Lekkerkerker}(1992)}]{vroege1992}
\bibinfo{author}{\bibfnamefont{G.~J.} \bibnamefont{Vroege}} \bibnamefont{and}
  \bibinfo{author}{\bibfnamefont{H.~N.~W.} \bibnamefont{Lekkerkerker}},
  \bibinfo{journal}{Rep. Prog. Phys.} \textbf{\bibinfo{volume}{55}},
  \bibinfo{pages}{1241} (\bibinfo{year}{1992}).

\bibitem[{\citenamefont{Straley}(1971)}]{straley1971}
\bibinfo{author}{\bibfnamefont{J.~P.} \bibnamefont{Straley}},
  \bibinfo{journal}{Phys. Rev. A} \textbf{\bibinfo{volume}{4}},
  \bibinfo{pages}{675} (\bibinfo{year}{1971}).

\bibitem[{\citenamefont{Frenkel and Eppenga}(1985)}]{frenkel1985}
\bibinfo{author}{\bibfnamefont{D.}~\bibnamefont{Frenkel}} \bibnamefont{and}
  \bibinfo{author}{\bibfnamefont{R.}~\bibnamefont{Eppenga}},
  \bibinfo{journal}{Phys. Rev. A} \textbf{\bibinfo{volume}{31}},
  \bibinfo{pages}{1776} (\bibinfo{year}{1985}).

\bibitem[{\citenamefont{Khandkar and Barma}(2005)}]{khandkar2005}
\bibinfo{author}{\bibfnamefont{M.~D.} \bibnamefont{Khandkar}} \bibnamefont{and}
  \bibinfo{author}{\bibfnamefont{M.}~\bibnamefont{Barma}},
  \bibinfo{journal}{Phys. Rev. E} \textbf{\bibinfo{volume}{72}},
  \bibinfo{pages}{051717} (\bibinfo{year}{2005}).

\bibitem[{\citenamefont{Vink}(2009)}]{vink2009}
\bibinfo{author}{\bibfnamefont{R.~L.~C.} \bibnamefont{Vink}},
  \bibinfo{journal}{Euro. Phys. J. B} \textbf{\bibinfo{volume}{72}},
  \bibinfo{pages}{225} (\bibinfo{year}{2009}).

\bibitem[{\citenamefont{Heilmann and Lieb}(1972)}]{lieb1972}
\bibinfo{author}{\bibfnamefont{O.~J.} \bibnamefont{Heilmann}} \bibnamefont{and}
  \bibinfo{author}{\bibfnamefont{E.}~\bibnamefont{Lieb}},
  \bibinfo{journal}{Commun. Math. Phys.} \textbf{\bibinfo{volume}{25}},
  \bibinfo{pages}{190} (\bibinfo{year}{1972}).

\bibitem[{\citenamefont{Dickman}(2012)}]{dickman2012}
\bibinfo{author}{\bibfnamefont{R.}~\bibnamefont{Dickman}}, \bibinfo{journal}{J.
  Chem. Phys.} \textbf{\bibinfo{volume}{136}}, \bibinfo{pages}{174105}
  (\bibinfo{year}{2012}).

\bibitem[{\citenamefont{de~Gennes and Prost}(1995)}]{degennesBook}
\bibinfo{author}{\bibfnamefont{P.~G.} \bibnamefont{de~Gennes}}
  \bibnamefont{and} \bibinfo{author}{\bibfnamefont{J.}~\bibnamefont{Prost}},
  \emph{\bibinfo{title}{The Physics of Liquid Crystals}}
  (\bibinfo{publisher}{Oxford University Press}, \bibinfo{address}{Oxford},
  \bibinfo{year}{1995}).

\bibitem[{\citenamefont{Ghosh and Dhar}(2007)}]{ghosh2007}
\bibinfo{author}{\bibfnamefont{A.}~\bibnamefont{Ghosh}} \bibnamefont{and}
  \bibinfo{author}{\bibfnamefont{D.}~\bibnamefont{Dhar}},
  \bibinfo{journal}{Euro. Phys. Lett.} \textbf{\bibinfo{volume}{78}},
  \bibinfo{pages}{20003} (\bibinfo{year}{2007}).

\bibitem[{\citenamefont{Disertori and Giuliani}(2012)}]{giuliani2012}
\bibinfo{author}{\bibfnamefont{M.}~\bibnamefont{Disertori}} \bibnamefont{and}
  \bibinfo{author}{\bibfnamefont{A.}~\bibnamefont{Giuliani}},
  \bibinfo{journal}{arXiv:1112.5564}  (\bibinfo{year}{2012}).

\bibitem[{\citenamefont{Matoz-Fernandez
  et~al.}(2008{\natexlab{a}})\citenamefont{Matoz-Fernandez, Linares, and
  Ramirez-Pastor}}]{fernandez2008a}
\bibinfo{author}{\bibfnamefont{D.~A.} \bibnamefont{Matoz-Fernandez}},
  \bibinfo{author}{\bibfnamefont{D.~H.} \bibnamefont{Linares}},
  \bibnamefont{and} \bibinfo{author}{\bibfnamefont{A.~J.}
  \bibnamefont{Ramirez-Pastor}}, \bibinfo{journal}{Euro. Phys. Lett}
  \textbf{\bibinfo{volume}{82}}, \bibinfo{pages}{50007}
  (\bibinfo{year}{2008}{\natexlab{a}}).

\bibitem[{\citenamefont{Matoz-Fernandez
  et~al.}(2008{\natexlab{b}})\citenamefont{Matoz-Fernandez, Linares, and
  Ramirez-Pastor}}]{fernandez2008b}
\bibinfo{author}{\bibfnamefont{D.~A.} \bibnamefont{Matoz-Fernandez}},
  \bibinfo{author}{\bibfnamefont{D.~H.} \bibnamefont{Linares}},
  \bibnamefont{and} \bibinfo{author}{\bibfnamefont{A.~J.}
  \bibnamefont{Ramirez-Pastor}}, \bibinfo{journal}{Physica A}
  \textbf{\bibinfo{volume}{387}}, \bibinfo{pages}{6513}
  (\bibinfo{year}{2008}{\natexlab{b}}).

\bibitem[{\citenamefont{Matoz-Fernandez
  et~al.}(2008{\natexlab{c}})\citenamefont{Matoz-Fernandez, Linares, and
  Ramirez-Pastor}}]{fernandez2008c}
\bibinfo{author}{\bibfnamefont{D.~A.} \bibnamefont{Matoz-Fernandez}},
  \bibinfo{author}{\bibfnamefont{D.~H.} \bibnamefont{Linares}},
  \bibnamefont{and} \bibinfo{author}{\bibfnamefont{A.~J.}
  \bibnamefont{Ramirez-Pastor}}, \bibinfo{journal}{J. Chem. Phys.}
  \textbf{\bibinfo{volume}{128}}, \bibinfo{pages}{214902}
  (\bibinfo{year}{2008}{\natexlab{c}}).

\bibitem[{\citenamefont{Linares et~al.}(2008)\citenamefont{Linares, Rom\'{a},
  and Ramirez-Pastor}}]{linares2008}
\bibinfo{author}{\bibfnamefont{D.~H.} \bibnamefont{Linares}},
  \bibinfo{author}{\bibfnamefont{F.}~\bibnamefont{Rom\'{a}}}, \bibnamefont{and}
  \bibinfo{author}{\bibfnamefont{A.~J.} \bibnamefont{Ramirez-Pastor}},
  \bibinfo{journal}{J. Stat. Mech.} p. \bibinfo{pages}{P03013}
  (\bibinfo{year}{2008}).

\bibitem[{\citenamefont{Fischer and Vink}(2009)}]{fischer2009}
\bibinfo{author}{\bibfnamefont{T.}~\bibnamefont{Fischer}} \bibnamefont{and}
  \bibinfo{author}{\bibfnamefont{R.~L.~C.} \bibnamefont{Vink}},
  \bibinfo{journal}{Euro. Phys. Lett.} \textbf{\bibinfo{volume}{85}},
  \bibinfo{pages}{56003} (\bibinfo{year}{2009}).

\bibitem[{\citenamefont{Barnes et~al.}(2009)\citenamefont{Barnes, Siderius, and
  Gelb}}]{barnes2009}
\bibinfo{author}{\bibfnamefont{B.~C.} \bibnamefont{Barnes}},
  \bibinfo{author}{\bibfnamefont{D.~W.} \bibnamefont{Siderius}},
  \bibnamefont{and} \bibinfo{author}{\bibfnamefont{L.~D.} \bibnamefont{Gelb}},
  \bibinfo{journal}{Langmuir} \textbf{\bibinfo{volume}{25}},
  \bibinfo{pages}{6702} (\bibinfo{year}{2009}).

\bibitem[{\citenamefont{Dhar et~al.}(2011)\citenamefont{Dhar, Rajesh, and
  Stilck}}]{dhar2011}
\bibinfo{author}{\bibfnamefont{D.}~\bibnamefont{Dhar}},
  \bibinfo{author}{\bibfnamefont{R.}~\bibnamefont{Rajesh}}, \bibnamefont{and}
  \bibinfo{author}{\bibfnamefont{J.~F.} \bibnamefont{Stilck}},
  \bibinfo{journal}{Phys. Rev. E} \textbf{\bibinfo{volume}{84}},
  \bibinfo{pages}{011140} (\bibinfo{year}{2011}).

\bibitem[{\citenamefont{Kundu et~al.}(2012)\citenamefont{Kundu, Rajesh, Dhar,
  and Stilck}}]{joyjit_dae}
\bibinfo{author}{\bibfnamefont{J.}~\bibnamefont{Kundu}},
  \bibinfo{author}{\bibfnamefont{R.}~\bibnamefont{Rajesh}},
  \bibinfo{author}{\bibfnamefont{D.}~\bibnamefont{Dhar}}, \bibnamefont{and}
  \bibinfo{author}{\bibfnamefont{J.~F.} \bibnamefont{Stilck}},
  \bibinfo{journal}{AIP Conf. Proc.} \textbf{\bibinfo{volume}{1447}},
  \bibinfo{pages}{113} (\bibinfo{year}{2012}).

\bibitem[{\citenamefont{Ramos et~al.}(1999)\citenamefont{Ramos, Rikvold, and
  Novotny}}]{Ramos1999}
\bibinfo{author}{\bibfnamefont{R.~A.} \bibnamefont{Ramos}},
  \bibinfo{author}{\bibfnamefont{P.~A.} \bibnamefont{Rikvold}},
  \bibnamefont{and} \bibinfo{author}{\bibfnamefont{M.~A.}
  \bibnamefont{Novotny}}, \bibinfo{journal}{Phys. Rev. B}
  \textbf{\bibinfo{volume}{59}}, \bibinfo{pages}{9053} (\bibinfo{year}{1999}).

\bibitem[{\citenamefont{Rikvold and Gorman}(1994)}]{rikvold1994}
\bibinfo{author}{\bibfnamefont{P.~A.} \bibnamefont{Rikvold}} \bibnamefont{and}
  \bibinfo{author}{\bibfnamefont{B.~M.} \bibnamefont{Gorman}},
  \bibinfo{journal}{Annual Reviews of Computational Physics}
  \textbf{\bibinfo{volume}{1}}, \bibinfo{pages}{149} (\bibinfo{year}{1994}).

\bibitem[{\citenamefont{Kenyon and Kenyon}(1992)}]{kenyon1992}
\bibinfo{author}{\bibfnamefont{C.}~\bibnamefont{Kenyon}} \bibnamefont{and}
  \bibinfo{author}{\bibfnamefont{R.}~\bibnamefont{Kenyon}}, in
  \emph{\bibinfo{booktitle}{Proc. of 33rd Fundamentals of Computer Science
  (FOCS)}} (\bibinfo{publisher}{IEEE Computer Society}, \bibinfo{address}{Los
  Alamitos, CA, USA}, \bibinfo{year}{1992}), pp. \bibinfo{pages}{610--619}.

\bibitem[{\citenamefont{Thurston}(1990)}]{thurston1990}
\bibinfo{author}{\bibfnamefont{W.~P.} \bibnamefont{Thurston}},
  \bibinfo{journal}{Amer. Math. Monthly} \textbf{\bibinfo{volume}{97}},
  \bibinfo{pages}{757} (\bibinfo{year}{1990}).

\bibitem[{\citenamefont{Ghosh et~al.}(2007)\citenamefont{Ghosh, Dhar, and
  Jacobsen}}]{ghosh2}
\bibinfo{author}{\bibfnamefont{A.}~\bibnamefont{Ghosh}},
  \bibinfo{author}{\bibfnamefont{D.}~\bibnamefont{Dhar}}, \bibnamefont{and}
  \bibinfo{author}{\bibfnamefont{J.~L.} \bibnamefont{Jacobsen}},
  \bibinfo{journal}{Phys. Rev. E} \textbf{\bibinfo{volume}{75}},
  \bibinfo{pages}{011115} (\bibinfo{year}{2007}).

\bibitem[{\citenamefont{Fortuin and Kasteleyn}(1972)}]{FK1972}
\bibinfo{author}{\bibfnamefont{C.~M.} \bibnamefont{Fortuin}} \bibnamefont{and}
  \bibinfo{author}{\bibfnamefont{P.~W.} \bibnamefont{Kasteleyn}},
  \bibinfo{journal}{Physica} \textbf{\bibinfo{volume}{57}},
  \bibinfo{pages}{536} (\bibinfo{year}{1972}).

\bibitem[{\citenamefont{Janke and Schakel}(2004)}]{Janke2004}
\bibinfo{author}{\bibfnamefont{W.}~\bibnamefont{Janke}} \bibnamefont{and}
  \bibinfo{author}{\bibfnamefont{A.~M.~J.} \bibnamefont{Schakel}},
  \bibinfo{journal}{Nuclear Physics B} \textbf{\bibinfo{volume}{700[FS]}},
  \bibinfo{pages}{385} (\bibinfo{year}{2004}).

\bibitem[{\citenamefont{Deng et~al.}(2004)\citenamefont{Deng, Bl\"{o}te, and
  Nienhuis}}]{Deng2004}
\bibinfo{author}{\bibfnamefont{Y.}~\bibnamefont{Deng}},
  \bibinfo{author}{\bibfnamefont{H.~W.~J.} \bibnamefont{Bl\"{o}te}},
  \bibnamefont{and} \bibinfo{author}{\bibfnamefont{B.}~\bibnamefont{Nienhuis}},
  \bibinfo{journal}{Phys. Rev. B} \textbf{\bibinfo{volume}{69}},
  \bibinfo{pages}{026123} (\bibinfo{year}{2004}).

\bibitem[{\citenamefont{Kamieniarz and Bl\"{o}te}(1993)}]{blote1993}
\bibinfo{author}{\bibfnamefont{G.}~\bibnamefont{Kamieniarz}} \bibnamefont{and}
  \bibinfo{author}{\bibfnamefont{H.~W.~J.} \bibnamefont{Bl\"{o}te}},
  \bibinfo{journal}{J. Phys. A} \textbf{\bibinfo{volume}{26}},
  \bibinfo{pages}{201} (\bibinfo{year}{1993}).

\bibitem[{\citenamefont{Kundu et~al.}()\citenamefont{Kundu, Rajesh, Dhar, and
  Stilck}}]{kundu2013}
\bibinfo{author}{\bibfnamefont{J.}~\bibnamefont{Kundu}},
  \bibinfo{author}{\bibfnamefont{R.}~\bibnamefont{Rajesh}},
  \bibinfo{author}{\bibfnamefont{D.}~\bibnamefont{Dhar}}, \bibnamefont{and}
  \bibinfo{author}{\bibfnamefont{J.~F.} \bibnamefont{Stilck}},
  \bibinfo{note}{in preparation}.

\end{thebibliography}

\end{document}